\documentclass[onecolumn,preprint]{revtex4}
\usepackage{amssymb}
\usepackage{amsmath}
\usepackage[dvips]{graphicx}

\begin{document}

\title{Molecular dynamics simulations of oxide memory resistors (memristors)
}
\author{S. E. Savel'ev$^{1}$, A. S. Alexandrov$^{1,2}$, A. M. Bratkovsky$%
^{2} $, and R. Stanley Williams$^{2}$}
\affiliation{$^1$Department of Physics, Loughborough University, Loughborough LE11 3TU,
United Kingdom\\
$^2$Hewlett-Packard Laboratories, 1501 Page Mill Road, Palo Alto, California
94304}

\begin{abstract}
Reversible bipolar nano-switches that can be set and read electronically in
a solid-state two-terminal device are very promising for applications. We
have performed molecular-dynamics simulations that mimic systems with oxygen
vacancies interacting via realistic potentials and driven by an external
bias voltage. The competing short- and long-range interactions among charged
mobile vacancies lead to density fluctuations and short-range ordering,
while illustrating some aspects of observed experimental behavior, such as
memristor polarity inversion.
\end{abstract}

\pacs{71.38.-k, 74.40.+k, 72.15.Jf, 74.72.-h, 74.25.Fy}
\maketitle

\section{Introduction}

Nanoscale metal/oxide/metal two-terminal non-linear circuit elements and
switches would be extremely attractive for dense memory, logic,
neurocomputing etc if they could scale well in size, power, driving voltage,
and can perform without much fatigue in a repeatable and uniform (across
e.g. the memory matrix) manner. Some of these issues are not well understood
although the hysteretic behavior of such materials, especially thin films of
Transition Metal Oxides (TMO) like Ta$_{2}$O$_{5}$, Nb$_{2}$O$_{5}$, TiO$%
_{2} $, NiO, Cu$_{2}$O, and Group III and IV oxides (Al$_{2}$O$_{3}$, SiO$%
_{x})$, in metal-insulator-metal (MIM) vertical devices have been studied
for decades \cite%
{krey60,krey62,gibb64NiO,chop65Ta2O5,arg68TiO2,hick70Nb2O5,stone,oxley77}.
The diverse semiconductor\cite{ovsh68,blom94,gaas07}, polymer\cite%
{staf68poly,ball75poly}, and correlated oxide systems \cite%
{jpl90,blom94,asamitsu97,bedn00,bedn01} exhibited switching under electric
pulsing. In recent years, interest in various oxide-based systems switchable
by an electric pulse has grown quite dramatically leading to important
breakthroughs and exposing various fundamental problems related to
mechanisms of bipolar (driven by alternating positive and negative biasing
to close the I-V hysteresis loop) and unipolar (driven by the bias of one
polarity)\ switching behavior \cite{stan0,stan,wasAM09}. In particular, it
was realized that bipolar switches are analogous to the `memristor', a
fourth passive circuit element originally postulated by Leon Chua in 1971%
\cite{stan0}.

There are challenges in understanding and controlling the coupled electronic
and ionic kinetic phenomena that dominate the behavior of oxide switching
devices like Pt/TiO$_{2}$/Pt, which is an exemplar memristor (resistor with
memory)\cite{stan}. It has been demonstrated unambiguously that bipolar
switching involves changes to the electronic barrier at the Pt/TiO$_{2}$
interface due to the drift of positively charged oxygen vacancies under an
applied electric field\cite{stan}. Various direct measurements revealed
formation of localized conducting channels in TiO$_{2}$: pressure modulated
conductance microscopy \cite{joshG09,miaoG09}, conducting atomic force
microscopy (AFM) \cite{ruth09},\ scanning transmission x-ray microscopy \cite%
{strachan09,strachan10}, and in-situ transmission electron microscopy \cite%
{kwon10}. On the basis of these measurements, it became quite clear that the
vacancy drift towards the interface creates conducting channels that shunt,
or short-circuit, the electronic barrier and switch the device ON\cite{stan}%
. The drift of vacancies away from the interface breaks such channels,
recovering the electronic barrier to switch the junction OFF. More recently,
Kwon \emph{et al.} \cite{kwon10} have performed the direct cross-sectional
TEM studies of the unipolar resistive switching of TiO$_{2}$, revealing the
presence of nanoscale Magneli phase Ti$_{4}$O$_{7}$ conductive channels
following device turning on and imaging the filament region in TEM. A Ti$%
_{4} $O$_{7}$ phase was confirmed from the temperature dependence of the
conductance. Concurrently, Strachan \emph{et al.} \cite%
{strachan09,strachan10} investigated the bipolar mode of TiO$_{2}$
switching, using non-destructive spatially-resolved x-ray absorption and
electron diffraction that allows nanometer scale studies of the associated
chemical and structural changes of a functioning memristive device. They
have observed that electroforming of the device is accompanied by forming an
ordered Magneli phase channel within the initially deposited (amorphous) TiO$%
_{2}$ matrix.

The observation of a Magneli crystallite in a titanium dioxide matrix shows
that electroforming Pt/TiO$_{2}$/Pt memristive system is related to a
localized partial reduction of titanium dioxide and crystallization of a
metallic conducting channel Ti$_{4}$O$_{7}$. Inside a titanium dioxide
matrix, the Magneli phases are thermodynamically favored over a high
concentration of randomly distributed vacancies, and thus they can act as a
source or sink of vacancies in the matrix material depending on the
electrochemical potential within the device\cite{liborio09}. The application
of an electrical bias can control vacancy motion in and out of this
sub-oxide phase, modulating a transport barrier and leading to the dramatic
conductivity change. It is worth noting that hysteretic switching in some
perovskite oxides monitored by TEM did not show any indication of conducting
channel formation \cite{fujii10}. Therefore, two types of models are usually
considered (i)\ in-plane homogeneous and (ii) in-plane inhomogeneous changes
accompanying switching in the material\cite{stone}. We shall study the
kinetics of oxygen vacancies in a generic situation like TiO$_{2}$ by way of
molecular dynamics to visualize processes taking place at the atomic scale
that are reminiscent of actual device behavior and whose origin is in
competing vacancy-vacancy interactions.

Another important aspect is that the forming step for the channel formation
in systems like TiO$_{2}$, NiO, VO$_{2}$ is accompanied by local heating
that is witnessed by the local emergence of high-temperature phases and
observed by thermal microscopy. It is worth noting that bipolar switching is
apparently field-driven and is very different from unipolar switching that
is power-driven. As a result, the bipolar switches may consume much less
power than the unipolar switches. At the same time, local heating does
accompany and seems to be important in both processes. Thus, fresh TiO$_{2}$
samples have an amorphous layer of titanium dioxide. To make the device
switch in a repeatable manner, a \emph{forming step} of rather large voltage
pulse is usually required to create a vacancy-rich region (forming is not
required if e.g. a vacancy-rich layer is provided intentionally by
fabrication). During this process, the TiO$_{2}$ anatase phase forms near
the conducting channel, which nominally requires temperatures over 350$%
%TCIMACRO{\U{b0}}%
%BeginExpansion
{{}^\circ}%
%EndExpansion
$C \cite{strachan10}. Studying local thermal effects during switching in TiO$%
_{2}$ provides strong evidence for local heating \cite{borg09}. Unambiguous
evidence for local heating accompanying conducting channel formation comes
from third harmonic generation \cite{noh3w08}; low resistance states (LRSs)
in NiO showing \emph{unipolar} switching were strongly nonlinear with
variations in the resistance R as large as 60\%, which was most likely
caused by the Joule heating of conducting channels inside the films. One can
note at this point that since the \emph{unipolar} switching is power driven,
therefore, it may be problematic to use it because of power requirements.

Given the above, the microscopic understanding of the atomic-scale mechanism
and identification of the material changes within an insulating barrier
appears to be invaluable for controlling and improving the memristor
performance. Within the volume 10$\times $10$\times $2 nm$^{3}$ of
perspective nano-memristors the number of oxygen vacancies could be as small
as a thousand, so that the conventional statistical approach for dealing
with many-particle systems might fail. Many important phenomena such as
dynamic phase transitions, as well as competition between thermal
fluctuations and particle-particle interactions in stochastic transport, can
be investigated using the Molecular Dynamics (MD) simulations of the
Langevin equations describing the thermal diffusion and drift of individual
interacting particles \cite{sav}. Here we present some preliminary results
of such simulations for a model memristor.

\section{Modeling a memristor}

In a semiconducting titanium dioxide, oxygen vacancies play an important
role in the electrical conduction. A stoichiometric TiO$_{2}$ crystal is a
colorless and transparent insulator. However, heated in vacuum, it loses a
part of its oxygen, or is reduced, and becomes dark blue in color and
electrically conductive. At high temperatures about 1000 C ionic conduction
becomes observable originating from oxygen and titanium vacancies \cite{miy}
while at ambient and lower temperatures small polarons carry the current
\cite{alemot}. The reduced rutile shows an anomalous electrical property
under a DC electric field of several tens of V/cm \cite{miy}. When a DC
electric field was applied to the bulk sample of reduced TiO$_{2}$, a single
pulse of electric current was observed together with a constant background
current. This pulsed current has been attributed to migration of oxygen
vacancies in reduced TiO$_{2}$ under a DC electric field. The oxygen
migration is visually observable under a DC electric field. Namely, when a
DC electric field was applied to the sample after repeated change of its
polarity, the color of the sample around the negative electrode changed into
dark blue due to the higher density of oxygen vacancies, while the region
around the positive electrode lost color \cite{miy}. In order to explain
these features, Miyaoka \emph{et al.} \cite{miy} have proposed that the
oxygen vacancies form clusters under a voltage, which may create crystalline
phases different from rutile, namely Magneli phases, as recently observed in
nanoscale probes\cite{kwon10,strachan10}.

Based on the observations of oxygen vacancy migration and clustering in bulk
\cite{miy} and nanoscale \cite{kwon10,strachan10} samples of TiO$_{2}$
induced by an electric-field, we model a memristor as shown in Fig. \ref%
{memristor}. In our model, there is a reduced rutile thin layer TiO$_{2-x}$
near one of the metallic electrodes stabilized by the Coulomb mirror
potential. Vacancies from this layer can drift toward the opposite electrode
in an electric field.
\begin{figure}[tbp]
\begin{center}
\includegraphics[angle=-90,width=0.90\textwidth]{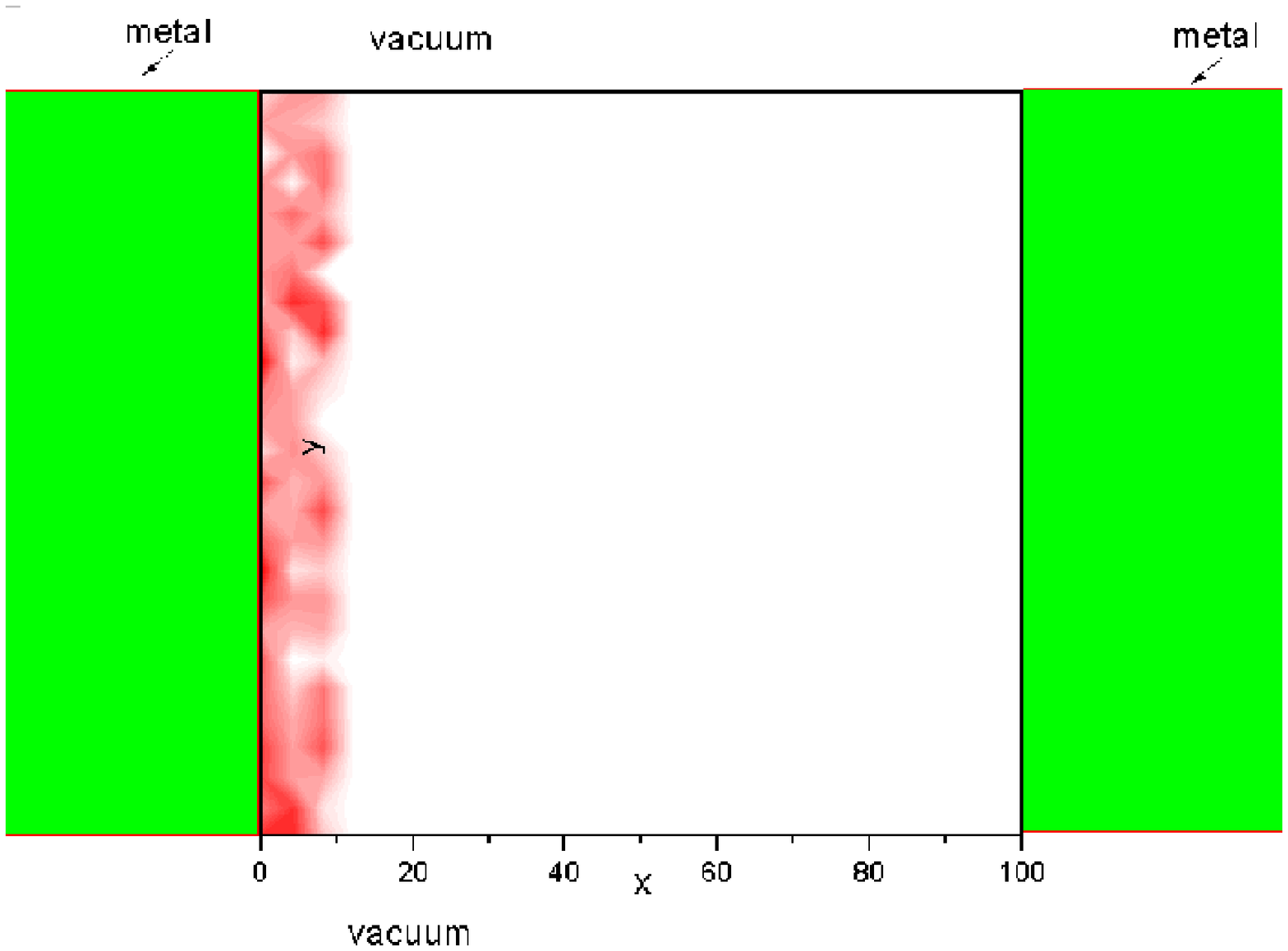}
\end{center}
\caption{(Color online) Model oxide memristor with a reduced oxygen layer
near one of the metallic leads (the length scale is arbitrary) }
\label{memristor}
\end{figure}

The simplest approach to the vacancy diffusion is using a drift-diffusion
equation for the density of vacancies $n(\boldsymbol{x},t)$,
\begin{equation}
{\frac{\partial n(\boldsymbol{x},t)}{{\partial t}}}=D\left[ {\frac{\partial
^{2}n(\boldsymbol{x},t)}{{\partial }\boldsymbol{x}{^{2}}}}-{\frac{%
\boldsymbol{F}(t)}{{k_{B}T}}}{\frac{\partial n(\boldsymbol{x},t)}{{\partial }%
\boldsymbol{x}}}\right] ,  \label{dif}
\end{equation}%
which neglects the particle-particle interaction. Here $D$ is the diffusion
coefficient, $T$ is the temperature, and $\boldsymbol{F}(t)$ is the electric
field. The diffusion equation (\ref{dif}) can be mapped onto a simple
Langevin equation
\begin{equation}
\frac{dx_{i}^{\alpha }}{dt}=F^{\alpha }(t)+\sqrt{D^{\alpha }}{\xi }%
_{i}^{\alpha }(t),
\end{equation}%
where $\alpha =x,y,z$ Cartesian components, ${\vec{\xi}}$ is a stochastic
force of zero mean and $\delta $-correlated in time, and we accounted for
anisotropic diffusion (in TiO$_{2}$ rutile diffusion along the c-axis is
much faster than in ab-plane, for instance).

\begin{figure}[tbp]
\begin{center}
\includegraphics[angle=-90,width=0.90\textwidth]{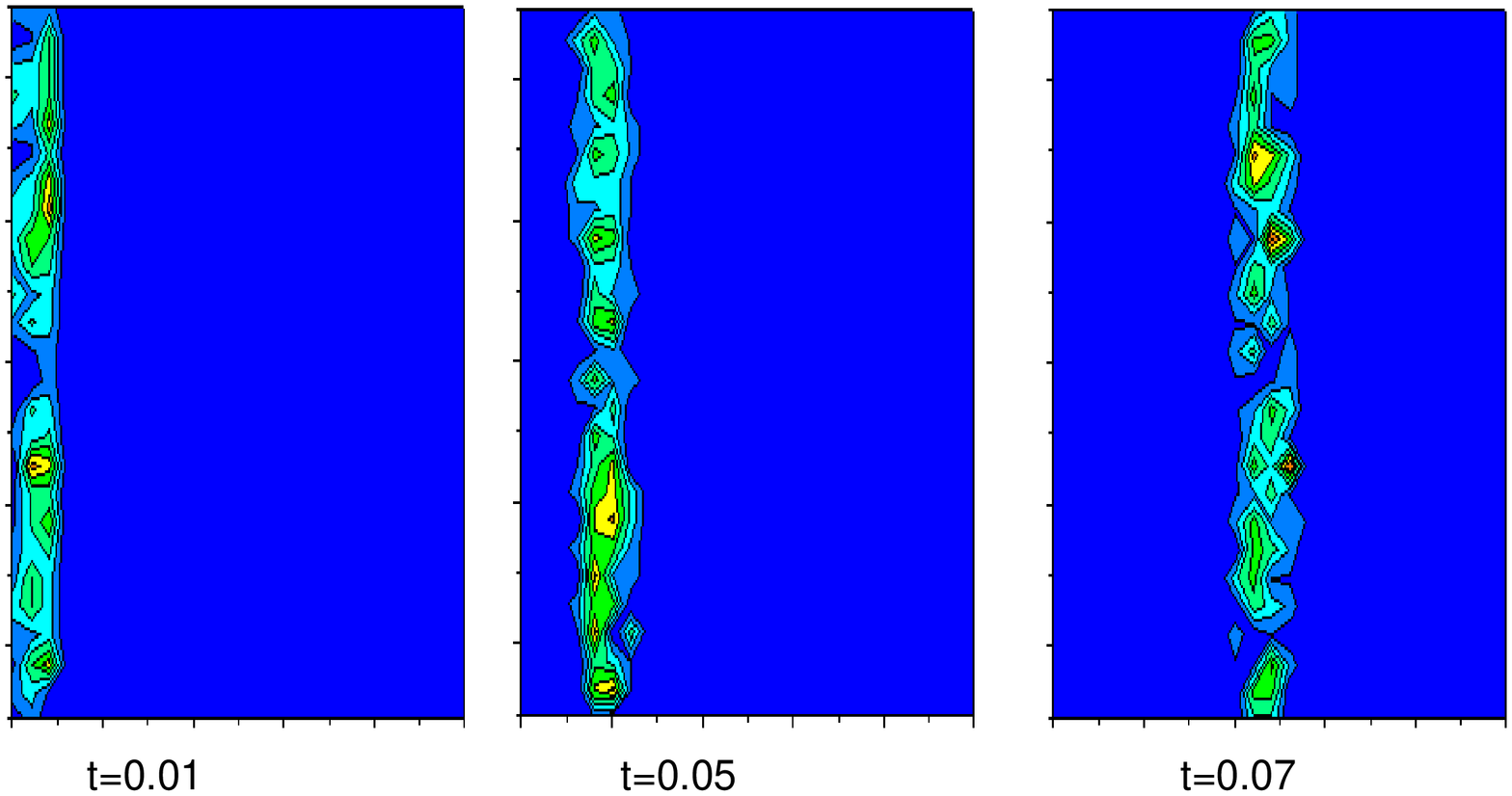}
\end{center}
\caption{(Color online) Simulated vacancy motion under the electric pulse
governed by the simplest diffusion equation (\protect\ref{dif}) showing no
clustering (length and time scales are arbitrary) }
\label{diffig}
\end{figure}

With the use of the standard Euler method\cite{sav}, we have simulated one
thousand vacancies ($i=1,...,1000$). The result is shown in Fig.\ref{diffig}
for a rectangular electric pulse, $F(t)$. There is no clustering in this
simple diffusion approximation, but rather uniform propagation of the
reduced layer accompanied by its widening in the propagation direction,
which is the same as the analytical result in this case.

\section{Molecular dynamics of interacting vacancies}

Now let us simulate the stochastic transport of vacancies fully taking into
account their interaction and boundary conditions. We consider an open
system of N point-like Brownian vacancies interacting with each other via
the pairwise potential $W$, with a possible inclusion of the substrate
through some periodic and random potentials $U$, and with internal and
external fields corresponding to the time-dependent deterministic force $F$.
The environment exerts an independent Gaussian random force, $\vec{\xi}$ on
each particle with zero mean and intensity controlled by the temperature. In
the overdamped regime (where inertia is negligible compared to the viscous
damping), the Langevin equation describing the drift-diffusion of the i-th
particle is \cite{sav}:
\begin{equation}
\eta {\frac{dx_{i}^{\alpha }}{{dt}}}=F_{i}^{\alpha }(\boldsymbol{x}%
_{i},t)-\sum_{j\neq i}{\frac{\partial W(\boldsymbol{x}_{i}\mathbf{-}%
\boldsymbol{x}_{j})}{{\partial }x{_{i}^{\alpha }}}}+\sqrt{2k_{B}T\eta }\xi
_{i}^{\alpha }(t).  \label{lan}
\end{equation}%
The unit of $\vec{\xi}_{i}$ is $s^{-1/2}$. The physical unit of length is
the thickness of the memristor, $L$, (about or less than $5$ nm). Then the
physical unit of time is $L^{2}/D$, where $D=k_{B}T/\eta $. (we take the
largest $D$ for anisotropic TiO$_{2}$). The diffusion coefficient $D$ should
be taken at a local temperature, but we have assumed some net temperature
across the active area in order to simplify matters in the present
simulation. Using the dimensionless time $\tau =tD/L^{2}$ and dimensionless
coordinates $\boldsymbol{r}_{i}=\boldsymbol{x}_{i}/L$, the equations (\ref%
{lan}) can be written as
\begin{equation}
{\frac{dr_{i}^{\alpha }}{{d\tau }}}=-{\frac{\partial \lbrack V(\boldsymbol{r}%
_{i},\tau )+U(\boldsymbol{r}_{i})]/(k_{B}T)}{{\partial }r{_{i}^{\alpha }}}}%
-\sum_{j\neq i}{\frac{\partial W(\boldsymbol{r}_{i}\mathbf{-}\boldsymbol{r}%
_{j})/(k_{B}T)}{{\partial }r{_{i}^{\alpha }}}}+\sqrt{2}\xi _{i}^{\alpha
}(\tau ),  \label{dimlan}
\end{equation}%
where $V(\boldsymbol{r}_{i},\tau )$ is the electric-field potential, and the
components of the (dimensionless) random force, $\xi _{i}^{\alpha }$ ($%
\alpha =x,y,z$), satisfies the fluctuation - dissipation relation $\langle
\xi _{i}^{\alpha }(0)\xi _{j}^{\beta }(\tau )\rangle =\delta (\tau )\delta
_{\alpha ,\beta }\delta _{i,j}$. Diffusion in TiO$_{2}$ is strongly
anisotropic, thus the diffusion along the c-axis and in-plane will be
described by two equations with different diffusion coefficients in more
accurate simulations.

The set of equations (\ref{dimlan}) is equivalent to an infinite BBGKY
(Bogoliubov - Born - Green - Kirkwood - Yvon hierarchy) set of equations for
multi-particle distribution functions. With no particle-particle interaction
it reduces to the drift-diffusion equation (\ref{dif}). Also, in a mean
field approximation, one obtains an additional drift term in the diffusion
equation, which however does not lead to clustering.

The O$^{2-}$ vacancy-vacancy interaction potential, which is crucial for
their clustering and phase transformations, can be modeled as
\begin{equation}
W(R)=A\exp (-R/a)-B/R^{6}+e^{2}/(\pi \epsilon _{0}\epsilon R),
\end{equation}%
where the short-range repulsive and attractive parts are represented with $%
A=22764.0$ eV, $a=0.01490$ nm, and $B=27.88\times 10^{-6}$ eV nm \cite%
{potential}, and the long-range Coulomb repulsion is the last term. More
generally the interaction parameters $A,B,a$ vary from one oxide to another,
and the dielectric constant $\epsilon $ may also vary from sample to sample.

To estimate the diffusion coefficient one can use an empirical equation of
Ref. \cite{rad},
\begin{equation}
D\left[ \text{cm}^{2}\text{/s}\right] =(1.03\times 10^{-3})\exp \left(
-E_{_{v}}/k_{B}T\right) ,
\end{equation}%
where the activation energy for oxygen vacancy diffusion has a poorly known
value, with estimates ranging from $0.5$ eV to about 1.1 eV. The lower
boundary, $E_{v}=0.5\mathrm{eV},$ corresponds to $D\approx 3.44\times
10^{-12}$ cm$^{2}$/s at the room temperature. For higher activation
energies, the vacancies would be immobile, as they are known to be at room
temperature. We do know, however, that they move during switching. In this
regard, it is worth noting that (i) another important parameter is the local
temperature during switching and (ii) strong local electric field
concentration at the tips of conducting channels/filaments. Some estimates
indicate that the local temperature may reach values like 650K, in which
case the diffusion coefficient would be much larger, obviously. The local
field concentration near tips of conducting channels/filaments is another
variable that will very strongly facilitate the growth of the channel and
needs to be simulated self-consistently along with local heating.

Given the uncertainties with ionic diffusion in memristors, at this stage we
would restrict ourselves to qualitative and semi-quantitative analysis. The
following parameters are chosen at this stage more for computational
convenience than to model actual devices. In actual devices the voltage drop
across the active region may be moderate, on the order of 1V, and the pulse
duration for switching may be just a few microseconds. With $L=5$ nm and
such diffusion coefficient the time unit is $L^{2}/D\approx 0.7$ ms. The
electric pulse rising time $100$~ns would correspond to $\tau _{r}=1.4\times
10^{-4}$ and an pulse length $10$ $\mu $s corresponds to $\tau
_{imp}=1.4\times 10^{-3}$. The pulse amplitude of $5$ V would then yield the
dimensionless force $F\equiv eV/(k_{B}T)\approx 200$ at room temperature.
The characteristic particle speed is $v=eV/L\eta $, so that the
dimensionless collision time is $\tau _{col}=a/LF\approx 1.5\times 10^{-5}$,
and the number of computing steps is $1/\tau _{col}\approx 70000,$
representing a challenge for MD simulations. The characteristic collision
distance $R_{c},$ where the collision force changes its sign from attraction
to repulsion, is in fact several times larger than $a$, which makes the
number of MD steps manageable. Also, decreasing the electric pulse amplitude
makes the number of the MD steps smaller. We will present simulations for
more realistic parameters elsewhere, with the inclusion of local thermal
effects directly into simulations.

\begin{figure}[tbp]
\begin{center}
\includegraphics[angle=-90,width=0.90\textwidth]{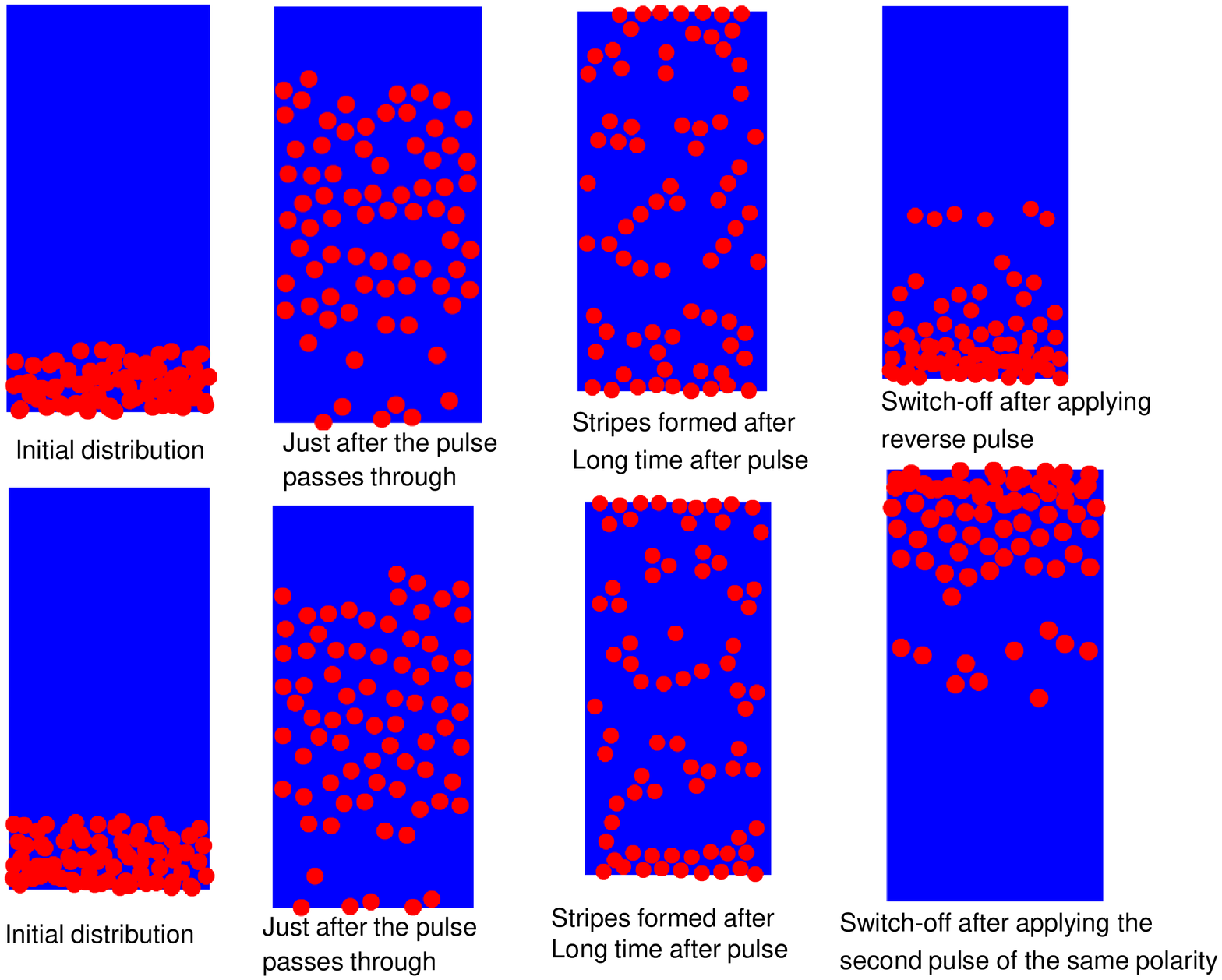}
\end{center}
\caption{(Color online) Two successive runs (upper and lower panels,
respectively) of the toy-model simulations with periodic BC and parameters
described in the text. The bottom panel illustrates the \textquotedblleft
polarity inversion" of the vacancies from bottom to top electrode during
successive pulsing. Analogous behavior has been observed experimentally in
Ref.\protect\cite{joshfam09}.}
\label{memristor3}
\end{figure}

\begin{figure}[tbp]
\begin{center}
\includegraphics[angle=-90,width=0.90\textwidth]{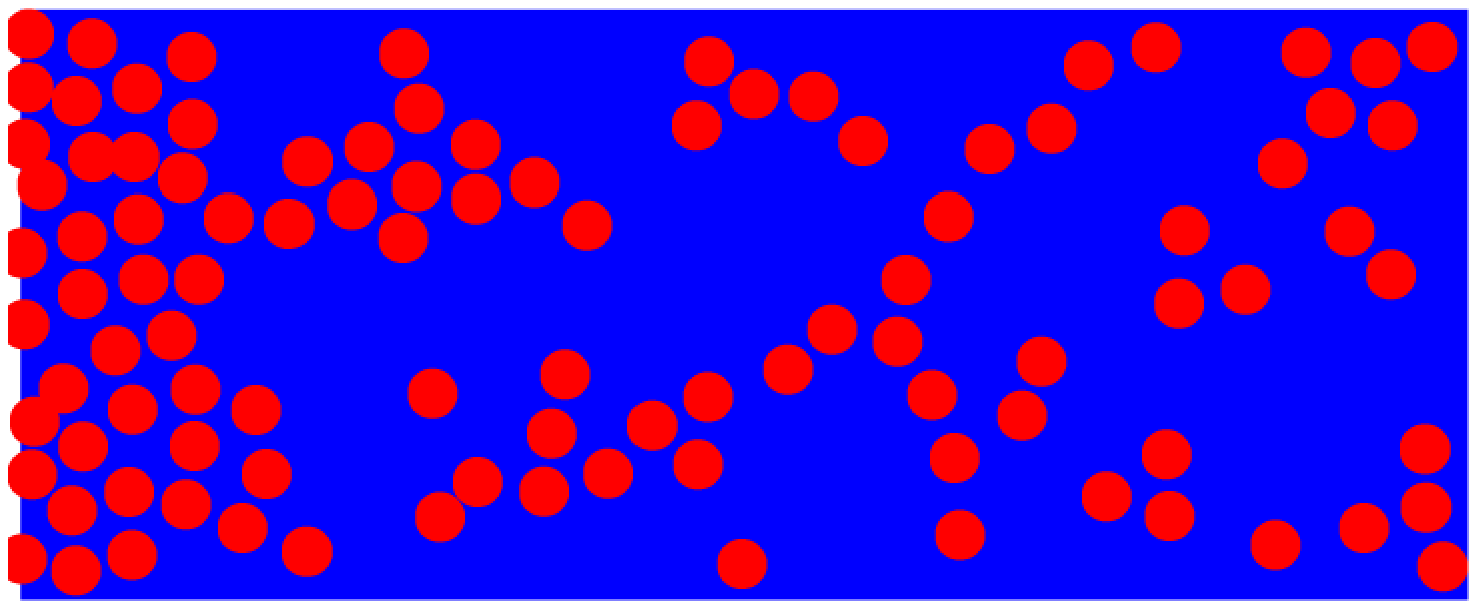}
\end{center}
\caption{(Color online) The same as in Fig.~\protect\ref{memristor3} with a
trend towards formation of a percolation path along the chains of oxygen
vacancies. }
\label{memristor4}
\end{figure}

\section{MD simulation of a toy memristor: clustering of vacancies}

In view of those uncertainties, we present here the MD simulations of a toy
memristor with relatively small number of vacancies, $N=75$ (Fig. \ref%
{memristor3}) and $150$ (Fig. \ref{memristor4}), leaving simulations of real
devices for future studies. Initially, we have randomly placed all the
vacancies near the bottom of a toy sample (see Fig. \ref{memristor3} for
initial distribution panels) and then let these vacancies evolve according
to Eq.(\ref{dimlan}) inside a rectangular box which mimics either a part
(Figs. \ref{memristor3}, \ref{memristor4}) or the entire (Fig. \ref%
{memristor5}) sample. We use the 2D\ simulation area with aspect ratio $%
L_{y}/L_{x}=2$ (Figs. \ref{memristor3},\ref{memristor4}) and $4/3$ (Fig. \ref%
{memristor5}) and either periodic boundary conditions (BC) (Fig.\ref%
{memristor3}, \ref{memristor4}) or impenetrable boundaries (Fig.\ref%
{memristor5}) along the x-direction. Note that periodic BC allow us to
simulate a quite large sample using a rather small number of particles,
while the impenetrable BC can elucidate the role of boundaries or 1D defects
in a real finite size sample. To simulate periodic BC, we include periodic
images of vacancies with respect to vertical boundaries of the simulation
box. We also incorporate opposite polarity charges by adding mirror images
of vacancies with respect to the top and the bottom of the sample.
\begin{figure}[tbp]
\begin{center}
\includegraphics[angle=-90,width=0.90\textwidth]{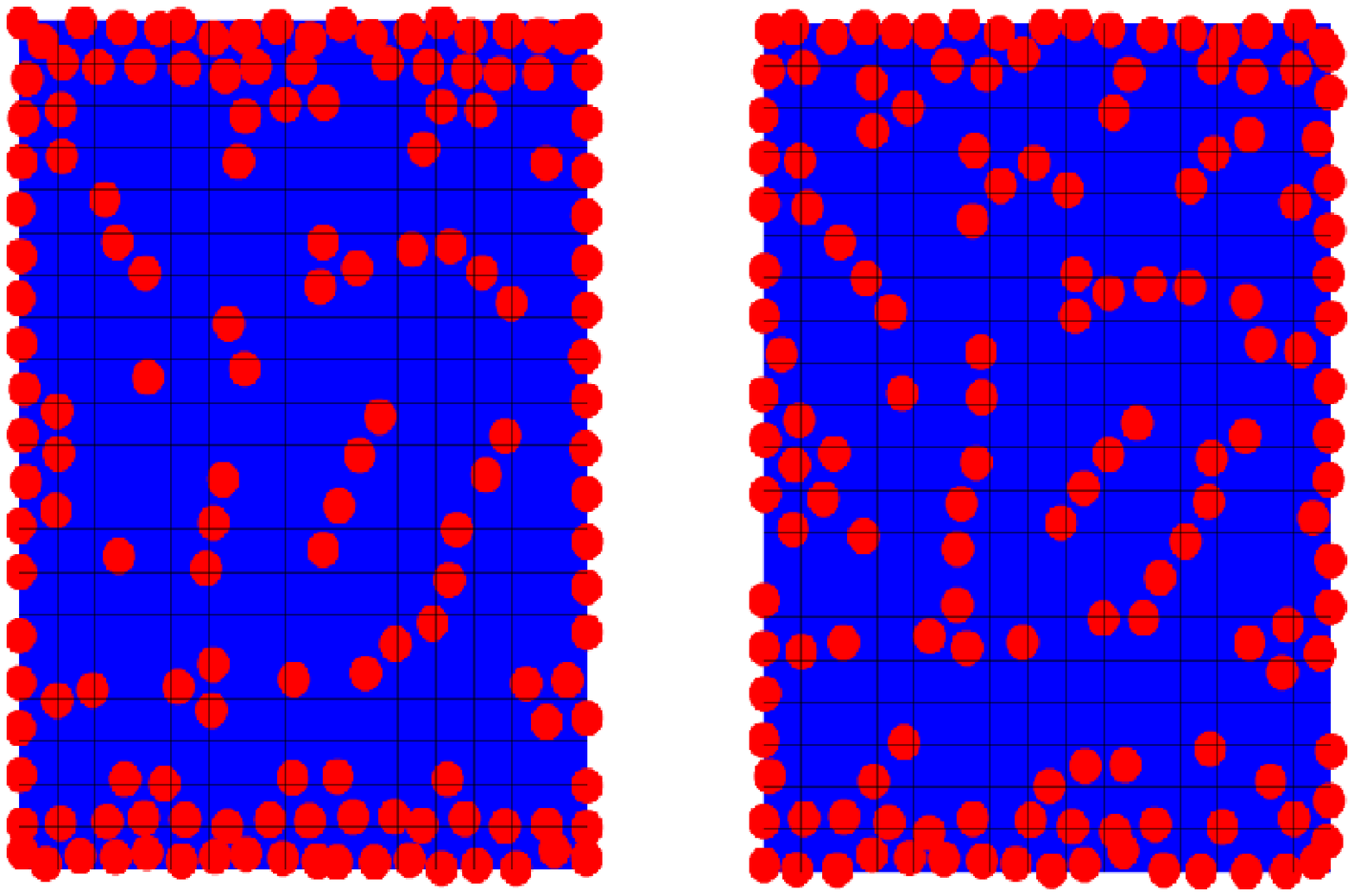}
\end{center}
\caption{(Color online) Results of toy-model simulations of a system with
impenetrable boundaries and parameters described in the text. As in Fig.~%
\protect\ref{memristor3}, one can see clustering of Oxygen vacancies.
Moreover, the sample boundary pins vacancies, favoring the formation of
percolation paths at the edges of the sample. The left and right panels
correspond to two successive runs with the same parameters but different
initial vacancy distributions. The observed edge effect is very reminiscent
of some memristors where switching areas tend to form near edges\protect\cite%
{strachan10}.}
\label{memristor5}
\end{figure}
Since we simulate a limited number of vacancies, one can refer to each
particle in our simulations as a cluster of vacancies, where a
cluster-cluster interaction described by the combination of the
Lennard-Jones and Coulomb potentials acting with the force
\begin{equation}
F(r)=\frac{1}{r}\left\{
12E_{LJ}[(r_{min}/r)^{12}-(r_{min}/r)^{6}]+E_{c}r_{min}/r\right\} ,
\end{equation}%
the relative strength of the Coulomb potential $E_{c}/E_{LJ}=2$ and the
potential well is about $30k_{B}$T. This results in the position of the
potential maxima $r_{max}\approx 2r_{min}$ and the height of the potential
barrier on the order of the depth of the potential well. The pulse strength
is taken to be about 10 times stronger than the maximum attracting force
between vacancy clusters.

The results of our toy-modeling look very promising. We see that pulsing in
opposite directions initially pushes the vacancies towards the top electrode
and then back, Fig.\ref{memristor3}. Pulsing twice in the forward direction
moves the vacancies from the bottom to the top electrode, thus inverting the
vacancy distribution in close analogy to what has been done experimentally%
\cite{joshfam09}. We observe a complex kinetics of the oxygen vacancies in
all present simulations. In particular, we have observed fragmentation of
vacancy distribution and formation of the short vacancy chains for both the
periodic and impenetrable boundary conditions. One can envisage that in
certain conditions those chains may form percolation paths (Fig. \ref%
{memristor4}), facilitated by e.g. pinning centers and/or boundaries of the
sample. Further studies would elucidate the conditions under which a
conducting channel may emerge in the sample driven by E-field, thus gaining
more insight into the switching of memristors.

\section{Conclusions}

We have proposed a model for the kinetic behavior of oxide memristors and
simulated its toy analog using the Molecular Dynamics Langevin equations.
Our MD simulations reveal a significant departure of the vacancy
distributions across the device from that expected within a standard
drift-diffusion approximation, driven by interactions among the mobile ionic
species (oxygen vacancies in TiO$_{2}$, as a generic example). A realistic
vacancy-vacancy interaction incorporating short and long-range repulsions
along with a medium-range attraction leads to clustering of vacancies into
chains. This may shed new light onto conducting channel formation in
memristors. We have found a significant effect of boundaries on the
clustering patterns. MD experiments (before averaging over configurations)
have the potential to simulate conditions that are hard or even impossible
to incorporate in the standard drift-diffusion models, such as the vacancy
annihilation or generation (from the air), different shapes of boundaries,
vacancy pinning (either periodic or random) and an inhomogeneous external
field (due to edge effects etc.). One can also calculate not only the
density distribution but binary distributions, velocity distributions etc.,
providing any desired statistical analysis of the distributions. Another
important step would be to include thermal effects via local changes in the
anisotropic diffusion coefficient. Competing interactions between mobile
ionic species are seen as a robust mechanism for vacancy concentration
fluctuations and clustering, which shows that switching in real devices is a
statistical or stochastic process.

%\begin{thebibliography}{99}

%\end{thebibliography}

\end{document}